\documentclass{jfm}
\usepackage{graphicx}
\usepackage{epstopdf, epsfig}
\usepackage{amsmath,makecell,hyperref,cleveref,url}

\hypersetup{
    colorlinks,
    linkcolor={blue},
    citecolor={blue},
    urlcolor={blue}
}

\newcommand\rar{\rightarrow}
\newcommand\bs{\boldsymbol}
\newcommand\ub{\mathbf{u}}
\newcommand\x{\mathbf{x}}

\newif\ifcomments
\newcommand\comments[1]{\ifcomments\textcolor{red}{#1}\else\relax\fi}
\commentsfalse

\newif\ifcommenttome
\newcommand\commenttome[1]{\ifcommenttome\textcolor{red}{#1}\else\relax\fi}
\commenttomefalse

\shorttitle{Prandtl-Batchelor theorem for quasi-periodic flows}
\shortauthor{H. Arbabi and I. Mezi\'{c}}

\title{Prandtl-Batchelor theorem for flows with quasi-periodic time dependence}

\author{Hassan Arbabi\aff{1}
  \corresp{\email{arbabi@mit.edu, mezic@engineering.ucsb.edu}},
Igor Mezi\'c\aff{2}}

\affiliation{\aff{1}Department of Mechanical Engineering, Massachusetts Institute of Technology
\\Cambridge, MA 02139, USA
\aff{2}Department of Mechanical Engineering, University of
California Santa Barbara \\Santa Barbara, CA 93106, USA }

\begin{document}

\maketitle

\begin{abstract}
The classical Prandtl-Batchelor theorem \citep{prandtl1904uber,batchelor1956steady}
states that in the regions of steady 2D flow where viscous forces are small and streamlines are closed, the vorticity is constant. In this paper, we extend this theorem to recirculating  flows with quasi-periodic time dependence using ergodic and geometric analysis of Lagrangian dynamics. In particular, we show that 2D quasi-periodic viscous flows, in the limit of zero viscosity, cannot converge to recirculating inviscid flows with non-uniform vorticity distribution. 
A corollary of this result is that if the vorticity contours form a family of closed curves in a quasi-periodic viscous flow, then at the limit of zero viscosity, vorticity is constant in the area enclosed by those curves at all times. 
\end{abstract}
\hrulefill

\section{Prandtl-Batchelor theorem}
The Prandtl-Batchelor theorem maintains that in the regions of steady flow where viscous forces are small and streamlines are closed, the vorticity is constant. 
The existence of such \emph{inviscid core} was first noted by \citet{prandtl1904uber}, and later formalized by \cite{batchelor1956steady}.  Batchelor also argued that the constant value of vorticity in the inviscid core can be determined by matching with the surrounding viscous boundary layer.
As the introduction to this theorem, we present its derivation for steady flows,  as stated by Batchelor, but we also include some modifications which makes the extension to time-dependent flows easier. 
The first step in proving this theorem is to derive a general condition on distribution of vorticity in the regions confined to closed streamlines. Consider the 2D Navier-Stokes equation,
\begin{equation}
  \frac{\partial \mathbf{u}}{\partial t}+\boldsymbol{\omega}\times \mathbf u +\nabla(\frac{u^2}{2})=-\nabla(\frac{p}{\rho})+\nu \nabla^2\mathbf u. \label{eq:NS}
\end{equation}
in a region of the flow where there is a nested family of closed streamlines. By integrating this equation along a closed streamline, we get
\begin{equation}
  \oint\frac{\partial \mathbf u}{\partial t}\cdot\mathbf{ds}+\oint\boldsymbol{\omega}\times \mathbf u\cdot\mathbf{ds} +\oint\nabla(\frac{u^2}{2}+\frac{p}{\rho})\cdot\mathbf{ds}=\oint\nu \nabla^2\mathbf{u}  \cdot \mathbf{ds}.
\end{equation}
In steady flows the leftmost term of the above vanishes identically. Moreover, $\mathbf{ds}$ is parallel to the velocity field and perpendicular to $\boldsymbol{\omega}\times \mathbf u$ at every point on the streamline, so the second term vanishes too. The last term on the left-hand side is also zero because it describes the change of a potential function over a closed curve. This leads to the conclusion that sum of the shear forces along the closed streamline is zero, i.e.,
\begin{equation}
\oint \nabla^2\mathbf{u}  \cdot \mathbf{ds}=\int_{C} \nabla^2\omega dA=0. \label{eq:NoShear2}
\end{equation}
where $C$ is the area enclosed by the streamline and we have used the Stokes theorem in the first equality. Using the divergence theorem we can write the above as
\begin{eqnarray}\label{eq:PBsteady}
\int_{C} \nabla^2\omega dA&=&\oint  \nabla \omega \cdot \mathbf{n}~ds=\oint \frac{\partial\omega}{\partial n} ~ds=0.
\end{eqnarray}
where $\partial\omega/\partial n$ is the outward derivative of vorticity in the direction normal to the streamline.

\comments{This paragraph is changed (assumption of regular limit):}
\comments{$$==============================================$$}
In the context of Prandtl-Batchelor theorem it is assumed that viscous effects are negligible such that vorticity is constant on each streamline.
Here, we use a slightly different form of this assumption, namely, we assume that the velocity field in the regions that contain the closed streamlines is given by $\ub=\ub_E+\ub_{\nu}$,  where $\ub_E$ is an inviscid solution, and $\ub_\nu$ is the viscous part of the velocity. Also we assume that the viscous part of velocity and the corresponding vorticity field vanish at the zero viscosity limit, that is,
\begin{equation}
\lim_{\nu\rar0} \ub_\nu=0, \quad \lim_{\nu\rar0}\omega_\nu=0.
\label{eq:reglim}
\end{equation}
Note that this notion of regular limit is slightly different (i.e. weaker) than the notion of regular limit used in the study of inviscid limit for viscous flows \citep[see e.g.][]{masmoudi2007remarks}. We will take this assumption throughout this paper.
Note that the this limit can be interpreted in two ways. It could be thought as the theoretical limit for dependence of solutions on viscosity. Or, similar to the setup by Batchelor, it can be considered as the limit of infinitely small viscous forces which is emulated at high Reynolds numbers and large distances from the walls.

\comments{$$==============================================$$}

Under the assumption of regular limit, when $\nu\rar0$, the vorticity is conserved along Lagrangian trajectories. In this situation, streamlines coincide with level sets of vorticity, since vorticity does not change for a tracer that circulates along the streamline.  Therefore, $\partial\omega/\partial n$ cannot change sign along a streamline, because as we move from a streamline toward outer streamlines, the vorticity either grows, or decays, or does not change - regardless of the position along that streamline (see \cref{fig:}).  Given that the last integral in \eqref{eq:PBsteady} is zero, the only possibility is that $\partial\omega/\partial n=0$ everywhere on the streamline which means that vorticity is also constant across the streamlines. Hence in all the region filled with these closed streamlines,
\begin{equation}
\omega=\omega_0~~const.
\end{equation}
Given that viscous effects decay with the increase of Reynolds number, and the assumption of regular limit holds, the Prandtl-Batchelor theorem predicted that vortices in steady flows at high Reynolds numbers and outside boundary layers would have nearly uniform vorticity in their core. This prediction was confirmed by experiments on the lid-driven flow in rectangular cavity \citep{pan1967steady}.

The Prandtl-Batchelor theorem has been extended in several directions:
\cite{grimshaw1969steady} generalized the theorem to weakly stratified flows and also showed that temperature behaves similar to vorticity when heat diffusivity is negligible.
\cite{blennerhassett1979three} and \citet{childress1990steady} have provided extensions of the theorem to 3D flows with helical symmetry. \citet{mezic2002extension} used the ergodic analysis of trajectories in steady 3D flows to obtain an integral condition on vorticity which reduces to \cref{eq:NoShear2} in the 2D setting. 
\citet{sandoval2010extension} also extended the theorem to 3D flows with slow variations in one direction. A few other extensions of this theorem (in Russian) are mentioned by \cite{sandoval2010extension}.

The caveat of the Prandtl-Batchelor theorem is that steady flows are highly unstable at high Reynolds numbers, and turn into flows with complex time dependence. To the best of our knowledge, however, there has been no extension of this theorem to time-dependent flows. In this paper, we prove this theorem for quasi-periodic flows with an arbitrary (but finite) number of basic frequencies, namely, we show that  in the limit of zero viscosity, the flow cannot converge to a recirculating inviscid flow with non-uniform vorticity distribution. The main concept that we employ in our proof is the notion of \emph{extended flow domain}, which is the extension of the flow domain in the time direction. 
A recirculating quasi-periodic flow is defined as a flow that possesses a nested family of invariant tori (i.e. nested tubes of Lagrangian path surfaces) in the extended flow domain,
By considering the ergodic properties of Lagrangian tracers, we derive a condition analogous to \eqref{eq:NoShear2}  which holds on the invariant surfaces in the extended flow domain. Then we use the geometry of these invariant surfaces to show that vorticity cannot change across them and therefore it is constant.

The outline of this paper is as follows: we first describe the ergodic aspects of vorticity evolution on tracers and assert the theorem for periodic flows in \cref{sec:periodic}. Then we extend the proof to quasi-periodic flows with an arbitrary number of basic frequencies in \cref{sec:qperiodic}. We present the numerical solutions of time-dependent lid-driven cavity flow---which confirms our theoretical results---in \cref{sec:cavity}. We summarize our discussion in \cref{sec:conclusion}.

\section{Periodic flow} \label{sec:periodic}
In what follows, we assume that the spatial flow domain $\mathcal{D}$ is a bounded subset of the plane. 
Central to our approach is the concept of \emph{extended} flow domain. Extended flow domain is the temporal extension of the spatial flow domain, with addition of time as an extra spatial direction. For a periodic flow, the time direction is periodic and the extended domain is the Cartesian product of the spatial domain $\mathcal{D}$ with the unit circle represented as $S=[0,2\pi)$. For example, if the spatial domain is a square, then the extended domain will be a square duct with periodic boundaries on two ends.  Also let $\Omega$ be the basic frequency of the flow. The motion  of tracers in the extended flow domain is then ruled by
\begin{eqnarray}\label{eq:pds}
  \left\{
  \begin{aligned}
    \dot{\mathbf{x}}&=\mathbf{u}(\theta,\x), \quad \x\in \mathcal{D}\\
    \dot{\theta}&=\Omega,\quad \theta\in S
  \end{aligned}
  \right.
\end{eqnarray}
The above system is a 3D time-\emph{independent} volume-preserving dynamical system. By the virtue of Birkhoff's ergodic theorem, the infinite-time averages of functions along the trajectories of this system exist and are finite for almost every initial condition \citep{petersen1989ergodic}. To formalize this, let $f(\theta,\x)$ denote the quantity of interest such as velocity, vorticity etc. Then the ergodic theorem asserts the existence of the limit
\begin{equation}
f^*:=\lim_{T\rar\infty}\frac{1}{T}\int_0^{T}fdt
\end{equation}
where the integration is done along a tracer trajectory.
Note that the material derivative of field quantities depend only on the position of the tracer in the extended domain (that is $\theta$ and $
\x$), and therefore, their time averages also exist and take finite values.
Now consider the infinite-time average of the material derivative of vorticity ($D\omega/Dt$) along a tracer trajectory. Due to the ergodic theorem, this time average not only exists, but it should be zero, otherwise vorticity would grow unboundedly on a tracer which is impossible in a periodic flow, i.e.,
 \begin{equation}
  \bigg( \frac{D\omega}{Dt} \bigg)^*=0 \label{eq:Dw-average}
\end{equation}
Now consider the 2D Navier-Stokes equation,
\begin{equation}
  \frac{D\omega}{Dt}=\nu\nabla^2\omega.
\end{equation}
By applying the infinite-time average along a tracer trajectory, and given that the left-hand side vanishes as in \eqref{eq:Dw-average}, we obtain
\begin{eqnarray}
0&=&\big(\nabla^2\omega\big)^*.\label{eq:lw0}
\end{eqnarray}

This observation, originally made by \citet{mezic2002extension}, holds for all viscous flows independent of the value of viscosity as long as $\nu\neq 0$. Moreover, under the assumption of regular limit in \eqref{eq:reglim}, it is also valid for $\nu=0$.

\comments{Major changes from here to the end of section
$$==============================================$$}

In the case of steady flow, it was assumed that the streamlines formed a nested family of closed curves. 
We generalize this for time-periodic flows, by assuming that there is a nested family of invariant tori (i.e. a family of tracer path surfaces in the form of nested tubes) in the extended domain of the inviscid flow. A tracer starting inside this family will rotate on the surface of an invariant torus and its motion will have a semicircular path in the spatial domain. 
Note that for inviscid flows, vorticity is an invariant of tracer motion (i.e. it is constant on a tracer path) and therefore its level sets coincide with those invariant tori.

\begin{rmk*}
The above recirculating assumption  is known to hold for inviscid flows with analytic velocity field and analytic boundary. In fact, it is shown by \citet{arnold1989mathematical} that the extended flow domain for such flows is filled (except on a zero-volume set) with nested families of invariant tori which are level sets of vorticity, and on each invariant torus, a tracer trajectory densely fills the surface. 
\end{rmk*}

We prove, by contradiction, that the inviscid flow described above cannot be the regular limit of a viscous flow, unless vorticity is constant across the invariant tori.

Let us denote an invariant torus of the inviscid flow described above with $T$, and the volume enclosed by it with $V_T$. Note that $V_T$ is also an invariant set of \eqref{eq:pds}. Since the tracer dynamics is volume preserving, $V_T$ can be partitioned into specific sets such that on each of those sets motion of tracers is ergodic (according to the so-called ergodic partition or ergodic decomposition theorem, see e.g. \citet{Mane:1987}). In the analytic inviscid flow, those ergodic sets would be the invariant tori that fill the volume inside $T$.
Ergodicity means that  on each of those sets, the time averages are equal to the spatial average, i.e.,
\begin{equation}
\int_{\mathbb{\xi}}\nabla^2\omega dv =\big(\nabla^2\omega\big)_\xi^*,\label{eq:lw1}
\end{equation}
where $\xi$ is such an ergodic set. Moreover, the integration over $V_T$ is an integration over all such ergodic sets \citep{Mezic:1994}. Now if this flow is the regular limit of a viscous quasi-periodic flow, then \eqref{eq:lw0} holds everywhere except on a zero-volume set, and therefore we have
\begin{equation}
\int_{V_T}\nabla^2\omega dv =0.\label{eq:lw2}
\end{equation}

\begin{figure} \centerline{\includegraphics[width=.75\textwidth]{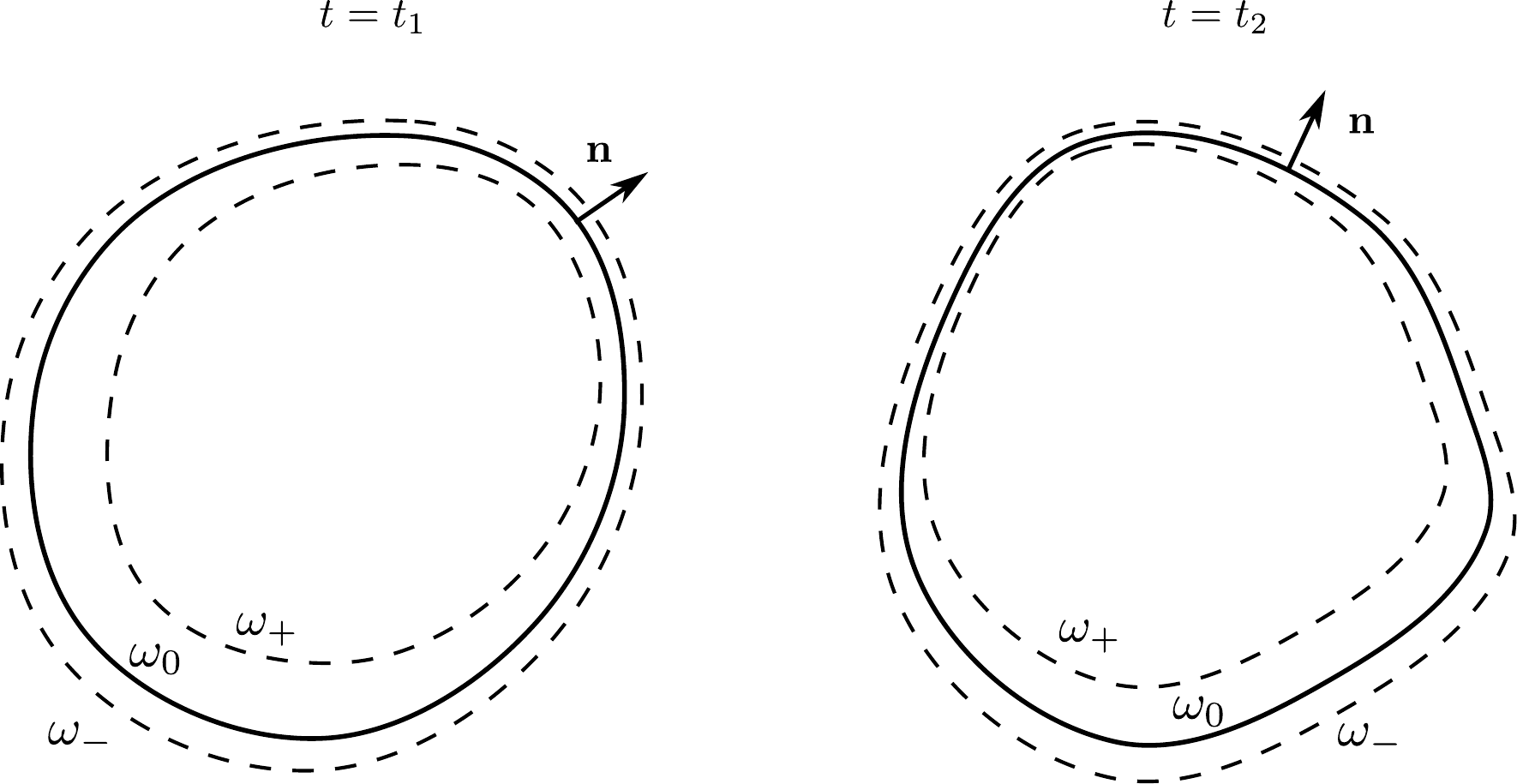}}
	\caption{The snapshot of vorticity contours in the inviscid flow at two different times. These curves are time-dependent barriers for advection and therefore cannot cross each other in time. As a result, $\partial\omega/\partial n$ does not change sign along the contour or in time. }
	\label{fig:}
\end{figure}

This observation will lead to contradiction unless vorticity is constant across the invariant tori. To see this, note that by Fubini's theorem, we can rewrite the above integral as
\begin{equation}
\int_{V_T}\nabla^2\omega dv =\int_{0}^{2\pi} \int_{S(\theta)} \nabla^2\omega ds~d\theta=0,
\end{equation}
where $S(\theta)$ is a $\theta$-slice of $V_T$ (i.e. intersection of $V_T$ with $\theta$-constant plane), or in fact, a closed curve of constant vorticity at time $t=\theta/\Omega$.
By applying the divergence theorem to the inner integral, we have
\begin{eqnarray}
\int_{V_T}\nabla^2\omega dv &=& \int_{0}^{2\pi} \int_{S(\theta)} \nabla^2\omega ds~d\theta=\int_{0}^{2\pi} \oint_{S(\theta)} \nabla\omega \cdot \mathbf{n} ds~d\theta \nonumber
\\ &=& \int_{0}^{2\pi} \oint_{S(\theta)} \frac{\partial\omega}{\partial n}(s,\theta) ds~d\theta=0 \label{eq:wkill}
\end{eqnarray}
where $s$ is  the coordinate along $S(\theta)$ and $\mathbf{n}$ is the normal to the curve $S(\theta)$ in the $\theta$-constant plane. Now note that as we cross a torus to another in the direction of $\mathbf{n}$, the vorticity either increases, decreases, or does not change, and therefore $\p \omega/\p n$ does not change sign over the torus (see \cref{fig:}). This implies that the $\p \omega/\p n$ is identically zero, and therefore the vorticity does not change across the tori. Given that vorticity is constant on each torus, this means that vorticity is constant inside the volume enclosed by the invariant tori.

\section{Quasi-periodic flow} \label{sec:qperiodic}
Consider the quasi-periodic flow
\begin{equation}
\ub_{q}(t,\x)=\ub(\Omega_1t,\ldots,\Omega_mt,\x), \label{eq:qpflow}
\end{equation}
where $\bs\Omega=[\Omega_1,\ldots,\Omega_m]$ are the $m$ basic frequencies of the flow which are rationally independent,  and $\ub$ is $2\pi$-periodic in its first $m$ arguments. Moreover, assume that function $\ub$ is twice continuously differentiable with respect to all its arguments. We define the extended flow domain $X$ to be the product of the spatial domain $\mathcal{D}$ with an $m$-torus, denoted by $\mathbb{T}^m=[0,2\pi)^m$, i.e., $X=\mathcal{D}\times \mathbb{T}^m$. We denote the spatial coordinates with $\x\in\mathcal{D}$, and periodic time coordinates  with  $\bs\theta=[\theta_1,\ldots,\theta_m]\in \mathbb{T}^m$.

The evolution of the flow field $\ub_q$ is defined only on a dense subset of $X$ given by
\begin{equation}
X_q=\cup_{t\geq0}(\mathcal{D},\bs\theta),\quad \bs{\theta}=\bs\Omega t\mod 2\pi.
\end{equation}
Now consider the vector field defined on $X_q$ given as
\begin{eqnarray}
\dot{\x}&=&\ub(\bs\theta,\x) \\
\dot{\theta}_i&=&\Omega_i,\quad (\x,\bs\theta)\in X_q\notag
\end{eqnarray}
which rules the motion of tracers in the quasi-periodic flow. A natural extension of this dynamical system to the whole $X$ is given by
\begin{eqnarray}\label{eq:qpds}
  \left\{
  \begin{aligned}
    \dot{\x}&=\ub(\bs\theta,\x)\\
   \dot{\theta}_i&=\Omega_i,\quad (\x,\bs\theta)\in X 
  \end{aligned}
  \right.
\end{eqnarray}
In the inviscid setting, vorticity is invariant on trajectories in the quasi-periodic flow in \eqref{eq:qpflow} which translates to
\begin{eqnarray}
\frac{D\omega}{Dt}=\frac{\p\omega}{\p t}+\ub\cdot\nabla \omega= \Omega_i\frac{\p\omega}{\p\theta_i}+\ub\cdot\nabla \omega=0,\quad (\x,\bs\theta)\in X_q,
\end{eqnarray}
where we have used the chain rule to the turn time derivative into spatial derivatives in the extended domain.
Now observe that $D\omega/Dt$ is a continuous function on $X$ and it is zero on $X_q$ which is dense in $X$. Then it must be zero on whole $X$, and therefore vorticity is an invariant of motion for the dynamical system in \eqref{eq:qpds}.  This implies that invariant sets of the flow in $X$, which are level sets of vorticity, are $m+1$-dimensional continuously differentiable manifolds \citep[see e.g.][]{lee2003smooth}.

The dynamical system in \eqref{eq:qpds} is volume-preserving, and similar to the time-periodic setting (see the discussion preceding \eqref{eq:lw0}), we can use the ergodic theorem to assert
\begin{eqnarray*}
0&=&\big(\nabla^2\omega\big)^*.\notag
\end{eqnarray*}

\comments{Major changes here:
$$==============================================$$}

Now we make the recirculation assumption similar to the time-periodic setting, that is, we assume that the invariant surfaces of inviscid flow form a nested family of tori in $X$. \comments{This assumption seems too strong and maybe it can be relaxed?}
Similar to periodic flows, we use proof by contradiction to show that such flow cannot be the regular limit of a quasi-periodic viscous flow unless vorticity is constant across the invariant surfaces.

Let $T$ denote an invariant torus described above. Then its enclosed volume, denoted with $V_T$, is also an invariant set of the dynamical system in \eqref{eq:qpds}. Under the assumption that such an inviscid flow is the regular limit of viscous flow
and using the argument preceding \eqref{eq:lw2}, we have
\begin{equation}
\int_{V_{T}}\nabla^2\omega dv =0.\label{eq:lw12}
\end{equation}

This statement again leads to contradiction unless vorticity is constant within $V_T$. Using Fubini's theorem, we can write the above as
\begin{eqnarray}
\int_{V_{T}}\nabla^2\omega dv &=&
\int_{0}^{2\pi}\ldots \int_{0}^{2\pi} \int_{S(\bs\theta)} \nabla^2\omega dA~d\theta_1\ldots d\theta_m \notag
\\
&=&\int_{0}^{2\pi}\ldots \int_{0}^{2\pi}\oint_{S(\theta,\phi)} \nabla\omega \cdot \mathbf{n} ds~d\theta_1\ldots d\theta_m		\notag
\\ &=& \int_{0}^{2\pi} \ldots \int_{0}^{2\pi} \oint_{S(\bs\theta)} \frac{\partial\omega}{\partial n}(s,\bs\theta) ds~d\theta_1\ldots d\theta_m=0 \label{eq:wkill2}
\end{eqnarray}
where $S(\bs\theta)$ is the $\bs\theta$-slice of $T$.
Note that by using the planar divergence theorem, we changed the domain of the integral from $V_T$ to $T$.

Now let $S(\bs\theta_1=\bs\Omega t_1)$ be a slice of $T$ which is also a closed contour of vorticity at $t=t_1$. Tow snapshots of this closed contour is sketched in \cref{fig:}. Similar to the steady flow, at each time, $\partial\omega/\partial n$ cannot not change sign over this closed curve.  Furthermore the sign cannot change with time either. To see this, note that this closed contour is a time-dependent advection barrier in the inviscid flow, i.e., the tracers inside it cannot travel outside, or vice versa, since it requires them to change the value of vorticity while vorticity is an invariant of motion. This means that outward derivative of vorticity on the curve does not change sign since local tracers with higher (res. lower) vorticity remain outside and the ones with lower (res. higher) vorticity stay inside. To make this precise let 
\begin{equation}\label{eq:M}
M:=\cup_{t\geq t_1}\big(S(\bs\Omega t),\bs\Omega t\big),
\end{equation}
be the time-extended image of the closed curve. 
The preceding statement says  that sign of $\partial\omega/\partial n$ does not change over $M$ . But $M$ is dense in $T$ and therefore the sign cannot change on $T$. Moreover $\p \omega / \p n$ cannot have a positive sign on $T$; if it is positive somewhere on $T$, noting that $\p\omega/\p n$ is a continuous function, the last integral in \eqref{eq:wkill2} would have a finite positive value which is a contradiction. The same  arguments rules out the negative sign for $\p\omega/\p n$, which leaves only $\p\omega/\p n=0$ everywhere on $T$. This means that the vorticity does not change along $\mathbf{n}$ at any time and therefore it must be constant in the volume that is enclosed by the family of invariant tori. 
The $\mathbf{\theta}$-slices of those tori at $\mathbf{\theta}=\mathbf{\Omega}t,~t\in \mathbb{R}$ are rotating invariant blobs in the quasi-periodic flow.
We summarize the results in the following theorem.

\begin{thm*} [Prandtl-Batchelor for quasi-periodic flows] 
Consider an incompressible inviscid flow in a bounded 2D domain with the quasi-periodic velocity field $$\ub_q(t,\x)=\ub(\Omega_1t,\ldots, \Omega_mt,\x)$$
where  $\Omega_i,~i=1,\ldots,m$ are the basic frequencies and $\ub$ is twice continuously differentiable with respect to all its arguments. Moreover, assume that the flow has a family of invariant surfaces in the form of $m+1$-dimensional nested tori in the extended flow domain.
Then a quasi-periodic viscous flow can converge to this flow at the regular limit of $\nu\rar 0$ (described in \eqref{eq:reglim}), \emph{only if} the vorticity is uniformly distributed across those surfaces.
\end{thm*}

We present an alternative formulation of the above result that describes a condition on the viscous flow that guarantees the existence of recirculating surfaces in the regular limit, namely, we assume that vorticity contours form a nested family of closed contours in the viscous flow at some instant. Note that this assumption does not hold for steady flows (though it is possible in unsteady flows), and therefore the following corollary is not the generalization of the classic theorem for the steady flow.

\begin{cor*} 
Consider an incompressible viscous flow in a bounded 2D domain with the quasi-periodic velocity field $$\ub_q(t,\x)=\ub(\Omega_1t,\ldots, \Omega_mt,\x)$$
where  $\Omega_i,~i=1,\ldots,m$ are the basic frequencies and $\ub$ is twice continuously differentiable with respect to all its arguments. 
Assume vorticity contours form a family of nested closed curves at some instant and this assumption holds at the regular limit of $\nu\rar 0$ described in \eqref{eq:reglim}. Then at this limit, vorticity is constant in the area enclosed by those closed curves at all times.
\end{cor*}

\begin{proof} From the assumption of vorticity contours forming a family of nested closed curves we can assert that the invariant manifolds in $X$ at the inviscid limit are a nested family of tori. To see this, recall that in the inviscid flow, vorticity is an invariant of motion and its level sets are $m+1$-dimensional manifolds in the extended flow domain. On the other hand a closed contour of vorticity in the viscous flow which persist in the limit is a $\theta$-slice of such level set. 
Let $T$ be a connected component of this level set. Then it follows from \citet[][Thm. 10]{susuki2018uniformly}  that $T$ must be a torus. Applying this to the family of closed curves yields that invariant surfaces in extended domain are a family of nested tori and the rest of the proof follows as before, i.e., assuming vorticity varies across invariant tori leads to contradiction. The $\mathbf{\theta}$-slices of those tori at $\mathbf{\theta}=\mathbf{\Omega}t,~t\in \mathbb{R}$ would be the area enclosed by the closed contours of  vorticity in the quasi-periodic flow.
\end{proof}

\comments{$$==============================================$$}

\section{Numerical example: 2D lid-driven cavity flow}\label{sec:cavity}
The lid-driven cavity flow is a widely used benchmark problem for evaluation of computational schemes for incompressible flows. This flow consists of a viscous fluid flow inside a box which is driven by the motion of one of the walls. \citet{burggraf1966analytical} used numerical simulations of cavity flow for Reynolds numbers up to 400 to verify the existence of the inviscid core predicted by the Prandtl-Batchelor theorem.  \citet{pan1967steady} also performed an experimental study of cavity flow with various aspect ratios and concluded that for very large Reynolds numbers the \emph{steady} flow will be dominated by a single rotating core of uniform vorticity. 

We investigate the Prandtl-Batchelor theorem in 2D lid-driven cavity flow by inspecting its numerical  solutions  for various Reynolds numbers.  The computational domain consists of the square box $\mathcal{D}=[-1,1]^2$ with stationary walls except the top lid which moves with the non-uniform velocity
\begin{equation}
u_\mathrm{top}=(1-x^2)^2,\quad x\in[-1,1],~y=1.
\end{equation}
This velocity profile is chosen to avoid singularities in the corners. By choosing the length of cavity side as the characteristic length scale, and the maximum velocity of the top lid as the characteristic velocity, we have defined the Reynolds number to be
\begin{equation}
Re= 2/\nu
\end{equation}
where $\nu$ is the kinematic viscosity in the numerical simulation.
We use the Chebyshev collocation scheme \citep{trefethen2000spectral}  to discretize the Navier-Stokes equations in space, and use a combination of Admas-Bashforth and Crank-Nicolson time stepping to integrate the resulting ordinary differential equations. 

\begin{figure} \centerline{\includegraphics[width=1.1\textwidth]{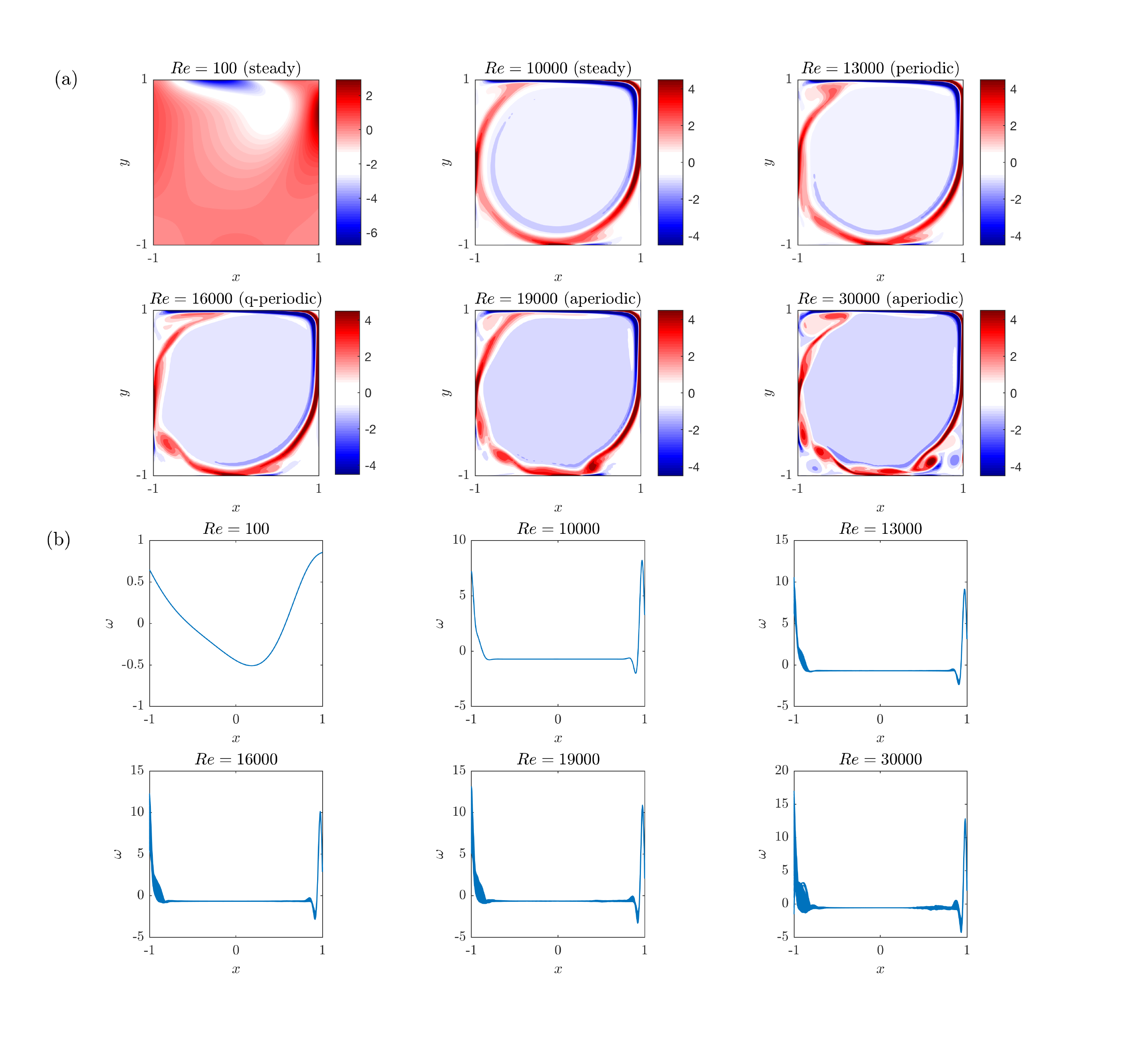}}
	\caption{(a) Snapshots of vorticity in the lid-driven cavity flow and, (b) the time trace of vorticity on the $y=0$-line over 100 simulation time units. }
	\label{fig:cavity}
\end{figure}

\Cref{fig:cavity} shows the snapshot of the vorticity field, as well as the trace of vorticity distribution on the line $y=0$, over 100 simulation time units of post-transient flow evolution.  
The cavity flow undergoes several bifurcations with the increase of Reynolds number  \citep{arbabi2017study}. The flow for Reynolds numbers up to 10000 converges to a steady solution. At high values of \Rey ~in this range, the flow consists of a central inviscid vortex surrounded by weak eddies in the downstream corners, i.e., all corners except the top right. At $Re\approx10500$, the flow undergoes a Hopf bifurcation and becomes time-periodic; the edge of the central vortex becomes unstable and hosts traveling waves that travel downstream along the edge. At $Re\approx15500$, the flow becomes quasi-periodic with two basic frequencies; the second frequency corresponds to another traveling wave along the edge of the core vortex. For high-Reynolds steady flows the central vortex of the cavity flow enjoys nearly uniform vorticity as predicted by classic theorem (e.g. $Re=10000$ in the figure). For periodic and quasi-periodic flows, the recirculating vortex in the center still maintains a nearly uniform distribution of vorticity despite its edge showing time variation due to the traveling waves (manifested as packets in \cref{fig:cavity}), therefore confirming the theorem derived above. 

As shown in the figure, the numerical simulations suggest that the Prandtl-Batchelor theory also holds for the stationary aperiodic flow with strong quasi-periodicity (e.g. \Rey=19000) as well as stationary chaotic flows at Reynolds numbers up to 30000.  Note that the theoretical analysis we presented above is not readily extendable to flows with aperiodic time dependence, however, measure-preserving  chaotic systems (e.g. statistically stationary turbulence) can be approximated with arbitrary accuracy, using quasi-periodic approximation of motion in the state space \citep[see e.g.][]{govindarajan2018approximation}. 
This suggests that this theorem may be extended to stationary 2D turbulent flows, and hence predicting vortices with nearly inviscid core in those flows at large Reynolds numbers.

\commenttome{this paragraph needs constant update}
\section{Conclusion}\label{sec:conclusion}
We extended the classical Prandtl-Batchelor theorem to 2D flows with quasi-periodic time dependence, namely, we showed that a  inviscid flow with recirculating invariant surfaces in the extended flow domain can be the regular limit of a viscous flow only if the vorticity is constant across those surfaces. 
 This extension was made through geometric and ergodic analysis of motion in the extended flow domain. 
Given the numerical evidence from lid-driven cavity flow, the assumption of regular inviscid limit, and the fact that stationary aperiodic flows can be approximated arbitrarily well using quasi-periodic flows, we speculate that this theorem holds for stationary aperiodic 2D flows as well. This implies that vortices in 2D stationary turbulence should possess nearly inviscid cores at high Reynolds numbers.
In future work, we will examine the role of Prandtl-Batchelor theorem in determining the structure of advective mixing in high Reynolds flows.

\section{Acknowledgements}
This work was partially supported by ONR grant  N00014-14-1-0633  and ARO-MURI grant W911NF-17-1-0306. We also thank the anonymous referees for their comments and suggestions.
\newpage

\bibliographystyle{jfm}
\bibliography{quasi-periodic Prandtl-Batchelor.bbl}

\end{document}